\documentclass[aps,10pt,showpacs,showkeys,twocolumn]{revtex4}
\usepackage[dvips]{graphicx}

\begin{document}
\title{Crystal field states in conducting magnetic materials:\\
NdAl$_2$, UPd$_2$Al$_3$ and YbRh$_2$Si$_2$$^\spadesuit$}
\author{R. J. Radwanski}
\homepage{http://www.css-physics.edu.pl}
\email{sfradwan@cyf-kr.edu.pl}
\affiliation{Center of Solid State Physics, S$^{nt}$Filip 5, 31-150 Krakow, Poland,\\
Institute of Physics, Pedagogical University, 30-084 Krakow, Poland}
\author{Z. Ropka}
\affiliation{Center of Solid State Physics, S$^{nt}$Filip 5,
31-150 Krakow, Poland}

\begin{abstract}
The existence of CEF states and the magnetism in conducting
compounds NdAl$_2$, UPd$_2$Al$_3$ and
  YbRh$_2$Si$_2$ is discussed. We point out that in metallic compounds localized crystal-field states coexist
  with conduction electrons. These compounds are physical realization
  of an anisotropic spin liquid, which in case of the atomic-like configuration with an odd number
  of f electrons, is unstable with respect to spin fluctuations for T--$>$0
  K. YbRh$_2$Si$_2$ is discussed as a compound with the Kramers doublet ground state with a very low
  magnetic ordering temperature being a reason for the quantum
  critical point phenomena at low T.

\end{abstract}
\maketitle\vspace {-1.0cm}

\section{Introduction and the aim}
\vspace {-0.4cm} NdAl$_2$, UPd$_2$Al$_3$ and YbRh$_2$Si$_2$ are
all metallic compounds. The aim of this paper is to study
crystal-field (CEF) effects in these metallic compounds owing to
the fact that still the concept and the use of the crystal-field
approach is questioned, if applied to metallic compounds. There
are opinions that obtained agreements with many experimental data
within the CEF-based approaches are accidental (for further
details of these reproaches see ArXiv/cond-mat/0504199). For us
the many-electron CEF approach is obvious and well founded in
solid-state physics. The many-electron CEF approach captures the
essential physics of transition-metal compounds. We think that
much of the localized/itinerant controversy is related to
misunderstandings due to a long-lasting lack of the open
scientific discussion that would allow for good, scientifically
clear, formulation of the subject of the controversy. The CEF is
a result of an inhomogeneous charge distribution in a solid. So,
it is everywhere. Another problem is its quantification. In a
crystal due to the translational symmetry and a high point
symmetry the CEF potential can be quantified by CEF coefficients
that reflect multipolar charge moments of the surroundings. A
transition-metal cation with the incomplete 4f/5f shell serves as
a probe of this CEF multipolar potential showing the discrete
splitting of the localized 4f/5f states. Here we will present
this discrete splitting of the localized 4f/5f states in NdAl$_2$,
UPd$_2$Al$_3$ and YbRh$_2$Si$_2$. The shown transition-metal
cations have in these compounds an odd number of 4f/5f electrons.
Thus, they have charge-formed Kramers doublet ground state. Due
to the presence of the orbital moment all of these compounds can
serve as example of the anisotropic spin lattice, here chosen in
a sequence of the lowering ordering temperature.

\section{Ferromagnet N$\textrm{d}$A$\textrm{l}_2$}
\vspace {-0.4cm}
 NdAl$_2$ crystallizes in the cubic MgCu$_2$
structure and orders ferromagnetically below 65 K [1]. Among
other papers in Ref. 2 it has been shown that its magnetic and
electronic properties are very well described as being
predominantly determined by the Nd$^{3+}$ ion with the 4f$^3$
configuration. In Ref. 2 the temperature dependence of the
magnetic moment and of the heat capacity have been calculated.
Fig. 1, similar to that shown in Ref. [2] - here we use the same
parameters, only we derive now more physical quantities, shows a
discrete electronic structure of 10 lowest localized states
originating from the lowest multiplet J=9/2. The 4f$^3$
configuration has in total 120 localized states. The localized f
electrons coexist with itinerant electrons originating from outer
3 Nd electrons and from outer Al electrons. Itinerant electrons
form a band and are responsible for the conductivity. Their
contribution to the magnetic moment or to the heat capacity, for
instance, is negligible in comparison to that of localized
electrons.
\begin{figure}[ht]
\begin{center}
\includegraphics[width = 6.5 cm]{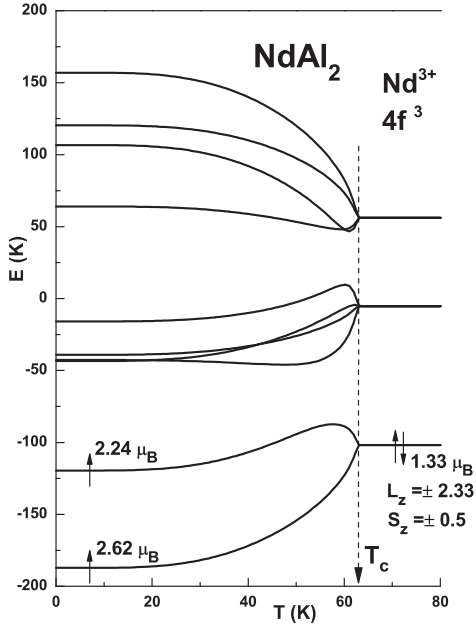}
\end{center}
\vspace {-0.8cm}\caption{Calculated temperature dependence of 10
states (J=9/2) of the ground multiplet of the Nd$^{3+}$ ion in
the paramagnetic and the magnetic state of the ferromagnet
NdAl$_2$. In a magnetic state, below 65 K, the Kramers degeneracy
is lifted and a spin gap is formed. The spin gap amounts at T = 0
K to 60 K. }
\end{figure}
The crystal-field splitting amounts to 160 K. Magnetic
interactions enlarge this splitting to 350 K. The calculated
magnetic moment at 0 K of 2.62 $\mu_B$ is in good agreement with
the experimental datum of 2.5$\pm$0.2 $\mu_B$ [1]. The calculated
total moment is composed from the spin moment m$_s$ = 3.35 $\mu_B$
and orbital moment m$_o$=-0.73 $\mu_B$. For calculations we make
use of a single-ion like Hamiltonian for the ground multiplet
J=9/2 [2]:
\\
\\
H~=~$H_{CF}$~+~$H_{f-f}$~=
\begin{equation}
  \label{eq:1}
  =~\sum \sum B_n^mO_n^m+n_{RR}g^2\mu _B^2\left(
-J\left\langle J\right\rangle +\frac 12\left\langle J\right\rangle
^2\right)\;.
\end{equation}

In Equation (1) the first term is the (cubic) crystal-field
Hamiltonian with B$_4$= -11.48 mK and B$_6$= +0.464 mK. These
parameters yield the multipolar moment of the Nd surroundings
A$_n^m$ = B$_n$/($\theta_n$ $\left\langle r_{4f}^n\right\rangle$
of A$_4$=+13.5 Ka$_o^{-4}$ and A$_6$=-0.81 Ka$_o^{-6}$. $\theta_n$
is a respective Stevens factor of the ground multiplet and
$\left\langle r_{4f}^n\right\rangle$ is the respective power of
the 4f radius. The second term takes into account intersite
spin-dependent interactions (n$_{RR}$ - molecular field
coefficient, magnetic moment $m$ = -$g$J$\mu_B$, $g$=8/11 - Lande
factor) - it produces the magnetic order below T$_c$ what is seen
in Fig. 1 as the appearance of the splitting of the Kramers
doublets and in experiment as the $\lambda$-peak in the heat
capacity at T$_{c}$.

\section{Heavy-fermion superconductor antiferromagnet UP$\textrm{d}_{2}$A$\textrm{l}_{3}$}
\vspace {-0.4cm}
 UPd$_2$Al$_3$ exhibits supercoductivity below 2
K coexisting with antiferromagnetism that appears below 14 K
[3,4]. A controversy is related to description of f electrons,
itinerant or localized in a number of 2 or 3. In our Refs. [5]
and [6], in contrary to authors of Refs [7,8] claiming the
5f$^{2}$ configuration, we have interpreted the excitations
observed by Krimmel {\it et al.} [9] as related to the energy
level scheme: 0, 7 meV (81 K), 10 meV (116 K) and 23.4 meV (271
K) of the $f^{3}$ (U$^{3+}$) configuration. The fine electronic
structure of the $f^{3}$ (U$^{3+}$) consists of five Kramers
doublets split by multipolar charge interactions (CEF
interactions). The fine electronic structure originates from the
lowest multiplet $^{4}$I$_{9/2}$, higher multiplets are at least
0.2 eV above and do not affect practically the ground-multiplet
properties. These 5 doublets are further split in the
antiferromagnetic state, i.e. below T$_N$ of 14 K as is shown in
Fig. 2. For calculations we make use the same Hamiltonian as for
the Nd ion in NdAl$_2$. A derived set of CEF parameters of the
hexagonal symmetry: B$_2^0$=+5.3 K, B$_4^0$=+40 mK, B$_6^0$=-0.02
mK and B$_6^6$=+26 mK yields states at 81 K, 120 K, 270 K and 460
K, energies of which are in perfect agreement with the
experimentally observed excitations. The highest state was not
observed in INS experiment. This electronic structure accounts
also surprisingly well for the overall temperature dependence of
the heat capacity, the substantial uranium magnetic moment and
its direction [6].

In Fig. 3 we present temperature dependence of the splitting
energy between two conjugate Kramers ground state. Surprisingly,
both the value of the energy and its temperature dependence is in
close agreement to a low-energy excitation of 1.7 meV at T=0 K
observed by Sato {\it et al.} [8] which has been attributed by
them to a magnetic exciton. We are convinced that the 5f$^{3}
$(U$^{3+}$) scheme provides a physical explanation for the 1.7
meV excitation (magnetic exciton) - this excitation is associated
to the removal of the Kramers-doublet ground state degeneracy in
the antiferromagnetic state.
\begin{figure}[ht]
\begin{center}
\includegraphics[width = 8.5 cm]{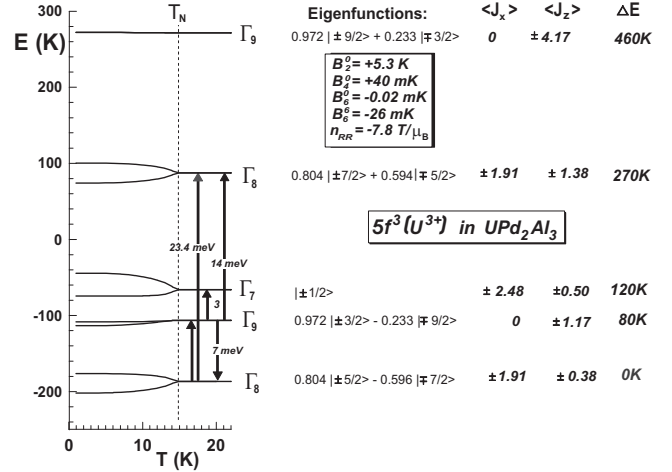}
\end{center}
\vspace {-0.5cm}\caption{Calculated energy level scheme of three
strongly-correlated f electrons, the 5f$^3$ configuration of the
U$^{3+}$ ion, in UPd$_2$Al$_3$. Arrows indicate transitions which
we have attributed to excitations observed by
inelastic-neutron-scattering experiments of Krimmel et al. [9].
After [6].}
\end{figure}

\begin{figure}[ht]
\begin{center}
\includegraphics[width = 7.2 cm]{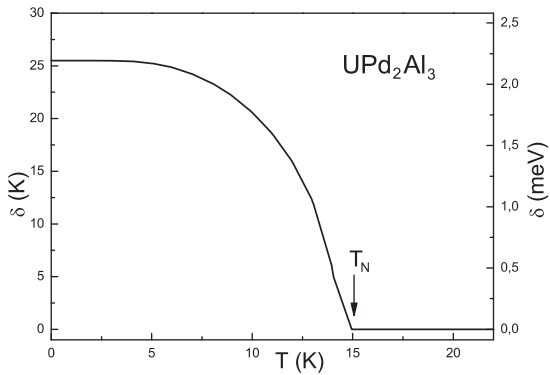}
\end{center}
\vspace {-0.5cm}\caption{Temperature dependence of the excitation
energy $\delta$ to the Kramers conjugate state of the $f^{3}$
(U$^{3+}$) configuration in UPd$_2$Al$_3$. This energy is
obtained as the difference between energies of two lowest states
seen in Fig. 2, taken from Ref. [6]. Both the value of the energy
and its temperature dependence is in close agreement to a
low-energy excitation of 1.7 meV at T = 0 K observed by Sato {\it
et al.} [8] and attributed by them to a magnetic exciton.}
\end{figure}

\section{Heavy-fermion metal Y$\textrm{b}$R$\textrm{h}_{2}$S$\textrm{i}_{2}$}
\vspace {-0.4cm}
 YbRh$_2$Si$_2$ is one of a hallmark
heavy-fermion compound with a Kondo temperature T$_K$ of 25-30 K.
It becomes recently of great scientific importance after the
discovery in Prof. F. Steglich group [10] of the well-defined
Electrons Spin Resonance (ESR) signal at temperature so low as
1.5 K similar to the ESR signal observed in conventional diluted
Yb alloys with the Yb$^{3+}$ ion. According to the Kondo model
below T$_K$ there should not be discrete states. The derived $g$
tensor is very anisotropic: $g_{\bot }$ = 3.561 and $g_{\Vert }$
=0.17 at 5 K. At 2 K $g_{\bot }$ decreases to 3.50. In Ref. [11]
we have derived two sets of CEF parameters that yield two Kramers
doublet ground-state eigenfunctions, $\Gamma _{6}^{1}$ or $\Gamma
_{7}^{1}$, that reproduce this g tensor at 5 K:

$\Gamma _{6}^{1}$ = 0.944 $|\pm 1/2>$ + 0.322 $|\mp 7/2>$ -

                              ~~~~~~~~~~~~~~~~~~~-0.052 $|\mp 3/2>$ - 0.047 $|\pm 5/2>$

$\Gamma _{7}^{1}$ = 0.803 $|\pm 3/2>$ + 0.595 $|\mp 5/2>$ -

                              ~~~~~~~~~~~~~~~~~~-0.027 $|\mp 1/2>$ - 0.009 $|\pm 7/2>$

In Ref. [12] we have proposed a physical experiment in order to
distinguish these two ground states. We have calculated
temperature dependence of the quadrupolar interactions for these
two states. As it is very different for these two ground states
the Mossbauer experiment should easily distinguish them. For
perfect reproduction of the $g$ tensor within the fully-localized
crystal-field model it was necessary to introduce the CEF B$_2^2$
term. It indicates lowering of the tetragonal symmetry of
YbRh$_2$Si$_2$ at low temperatures to the orthorhombic symmetry.
The increase of this distortion with lowering temperature is,
according to us, the reason for temperature variation of the $g$
tensor.

\section{Some remarks on the crystal-field approach}
\vspace {-0.4cm}
 A success in description of NdAl$_2$ within the
atomic-based CEF approach is related to a fact that there is only
one crystallographic Nd site up to lowest temperatures. Due to the
Kramers doublet ground state being not susceptible to the charge
surroundings and due to relatively large separation to the first
excited state mechanisms leading to a lowering symmetry are not
effective as by them the system cannot gain energy. NdAl$_2$
being ferromagnet assures conditions for all Nd moments to have
the same magnetic interactions. It causes that all Nd ions
equally contribute to the macroscopic magnetization and
thermodynamical properties. As a consequence, the macroscopic
molar values are obtained from the atomic-scale properties by
simple multiplication by the Avogadro number.

We advocate by years for the importance of the crystal-field
approach in 4f/5f compounds, also in those exhibiting
heavy-fermion phenomena. Of course, we do not claim that CEF
explains everything (surely CEF itself cannot explain the
formation of a magnetic ordering, for which in Eq. 1 we add the
second term), but surely the atomic-like 4f/5f/3d cation is the
source of the magnetism of a whole solid. The atomic-scale
magnetic moment is determined by local effects known as crystal
field and spin-orbit interactions as well as very strong electron
correlations. These strong electron correlations are
predominantly of the intra-atomic origin and taking into account,
in the first approximation, via on-site Hund's rules. These
strong correlations lead to many-electron version of the CEF
approach.

Finally, we use a name "$f$ electrons", as often is used in
literature. Of course, $f$ electrons are not special electrons but
it means electrons in $f$ states. $f$ states have well defined
characteristics, like the orbital quantum number $l$. By
identifying states in a solid we can identified $f$ electrons, in
particular their number.

\section{Conclusions}
\vspace {-0.4cm} Owing to a debate on the formation and a role of
localized crystal-field states in conducting magnetic materials
we point out that in metallic compounds localized crystal-field
states coexist with conduction electrons. The former are
predominantly responsible for magnetic and low-energy
spectroscopic properties, the latter for the conduction.

In NdAl$_2$ and UPd$_2$Al$_3$ we illustrate the physical
realization of the anisotropic spin liquid (lattice of Kramers
ions with an odd number of electrons), that is unstable with
respect to spin fluctuations for T--$>$0 K. In these three
compounds the magnetic ordering temperature decreases from 65 K
via 14 K to 0.07 K only in case of YbRh$_2$Si$_2$. The difficulty
in the removal of the Kramers degeneracy in case of
YbRh$_2$Si$_2$ is, according to us, the reason for the quantum
critical point phenomena observed at ultra-low temperatures.

In our atomic-start approach to 3d-/4f-/5f-atom containing
compounds we take into account crystal field and spin-orbit
interactions as well as very strong electron correlations,
predominantly of the intra-atomic origin.\\
$^\spadesuit$ Dedicated to the Pope John Paul II, a man of
freedom and truth in life and in Science.

\end{document}